\documentstyle[11pt,doublespace,fullpage]{article}

\begin{document}
\title{Cumulant Expansions and
 the Spin-Boson Problem\footnote{submitted to Phys. Rev. E}}
\author{David R. Reichman, Frank L. H. Brown, and Peter Neu \\
{\it Department of Chemistry} \\ 
{\it Massachusetts Institute of Technology} \\  
{\it Cambridge, Massachusetts, 02139}}
\maketitle
\thispagestyle{empty}
\begin{abstract}
The dynamics of the dissipative two-level system at zero temperature is
studied using three different cumulant expansion techniques.  The relative
merits and drawbacks of each technique are discussed.  It is found that
a new technique, the non-crossing cumulant expansion, appears to embody
the virtues of the more standard cumulant methods.\\ 

\noindent PACS-number(s): 05.30.-d, 05.40.+j, 72.15.Qm
\end{abstract}
\newpage

\setcounter{page}{1}

\section{Introduction}
The standard spin-boson problem, described by the Hamiltonian \cite{Legg,WB},
\begin{equation} \label{1}
H = \frac{\Delta}{2} \sigma_{x} + \sum_{j} \left[ \frac{p_{j}^{2}}{2m_{j}}
+ \frac{1}{2} m_{j} \omega_{j}^{2} \left(x_{j} -\frac{c_{j}}{m_{j}\omega_{j}^{2}}\,\sigma_z \right)^{2} \right],
\end{equation}
has served as a paradigm for the description of dissipative effects in condensed
phases.  Some experimental realizations of such a Hamiltonian include, e.g.,  the 
detection of macroscopic quantum coherence in superconducting quantum interference
devices \cite{Le,Esaki}, tunneling effects in metallic and insulating glasses \cite{P},
electron transfer reactions \cite{ET} and the diffusion of light
interstitial particles in metals \cite{int}.  In each situation, the physical realization of
the parameters in the Hamiltonian (1) is different.  For instance, in metallic
glasses at low temperatures the electron-hole pairs at the Fermi level constitute
the bosonic bath, while for insulating glasses, tunneling effects are damped
by localized and delocalized vibrational modes.  Thus, the Hamiltonian (1)
embodies a wealth of physical situations and has been studied in great detail
(see for instance \cite{Legg,WB} and references quoted therein, or
more recently \cite{KMN,KM,W}).

In order to study the dynamics of the two-level system coupled to a harmonic
bath as in (1), we need a method of ``tracing out'' the bath or spin degrees of freedom.
The bath  degrees of freedom can be specified by the spectral density function,
\begin{equation}\label{2}
J(\omega) =  \frac{\pi}{2} \sum_{j} \frac{c_{j}^{2}}{m_{j}\omega_{j}} \delta(\omega-\omega_{j}),
\end{equation}
which gives the bath density of states weighted by the square of the coupling
strength between the two-level system and the bath.  In most studies of the spin-boson problem, the spectral
density takes the ohmic form \cite{Legg,WB},
\begin{equation}\label{3}
J(\omega) = 2 \pi \alpha \omega \exp(-\omega/\omega_{c}),
\end{equation}
where $\alpha$ is a measure of the coupling strength, and $\omega_{c}$ is a frequency cutoff for the bath.
We note that in many cases, such as the coupling of a spin degree of freedom to a three dimensional phonon
bath in the deformation potential approximation, the spectral density (3) is not realistic, and must involve
higher powers of $\omega$.

The usual approach to finding reduced equations for the spin variables of interest involves the use of
the functional integral formulation of quantum dynamics \cite{Legg, WB}.
  Formally exact equations may be found for the 
variables
\begin{equation}\label{4}
P(t) = \langle \sigma_{z}(t) \rangle,
\end{equation}
and
\begin{equation}\label{5}
C(t) = \frac{1}{2} \langle \{ \sigma_{z}(0), \sigma_{z}(t) \} \rangle_{\beta} ,
\end{equation}
where $\langle ...\rangle_\beta$ refers to an average with respect to the canonical ensemble of (\ref{1}).
The quantity $P(t)$ describes the population difference in the localized spin 
states of the Hamiltonian (1), given that the particle is initially localized in one well 
and in thermal equilibrium with the bath.
It  is the variable of interest in certain physical situations, for example, the electron transfer problem
\cite{ET}.
The quantity $C(t)$, the symmetrized equilibrium correlation function of the tunneling coordinate, 
is related to the structure factor for neutron scattering off the tunneling particle, and
is of great significance in various problems, including the antiferromagnetic Kondo problem \cite{Aff}. 
 For $C(t)$ the long-time behavior at zero temperature is known from the generalized Shiba 
 relation which predicts algebraic decay $C(t) \propto t^{-2}$
 \cite{WB,Sas}.  For $P(t)$ the situation is less clear,
  however some studies have predicted exponential decay as $t \rightarrow \infty$ \cite{WB,Makri1}.
    Despite the
importance of $C(t)$, we will focus on the variable $P(t)$ in the following.

The formal path integral expression for $P(t)$ is extremely cumbersome, and a suitable approximation must be 
implemented to obtain useful information.  The so-called ``non-interacting blip approximation'', or NIBA
\cite{Legg,WB},
is the most commonly used approximation.  In this scheme $C(t)$ is entirely determined by $P(t)$, i.e.,
$C(t) \equiv P(t)$.
The NIBA may be obtained from the exact expression for $P(t)$ 
by invoking a series of physically based approximations.  For very low temperatures, these approximations
often break down, unless $\alpha$ is very small and only short times are considered.  At zero temperature,
the NIBA is not justified in the antiferromagnetic Kondo regime
 $\frac{1}{2} < \alpha < 1$.  The NIBA also incorrectly predicts
asymptotically algebraic, rather than exponential, decay for the variable $P(t)$.  Lastly, NIBA incorrectly
predicts that at zero temperature, $C(t) \sim t^{-2(1-\alpha)}$.  

Despite these flaws, the NIBA is  useful for obtaining quantitative results for $P(t)$ 
for high temperatures, when the tunneling dynamics is incoherent, and 
in predicting the qualitative behavior of $P(t)$ for low temperatures.  For instance, at zero temperature,
the NIBA correctly predicts a crossover from damped oscillations to incoherent decay for the variable $P(t)$
at the point $\alpha = \frac{1}{2}$.  

As shown by Aslangul et. al. \cite{As1},
 the NIBA may be obtained by first applying a small polaron transformation to the Hamiltonian (1),
followed by a second order application of the usual Nakijima-Zwanzig
 projection operator technique. It has been 
known for some time that this projection technique, which leads to a master equation of the convolution
form, is an order by order resummation of a particular type of cumulants known as ``chronologically ordered'' 
cumulants \cite{Ter,Deu,Mu1,Mu2,Shi}.  The use of the ``chronological ordering prescription'', or COP, when truncated at second order
thus leads to the NIBA.

Interestingly, Aslangul et. al. \cite{As2} 
earlier applied a convolutionless master equation technique to the study
of the zero temperature spin-boson problem.  This type of master equation, which can be derived by
using a different type of projection operator, involves the summation of a different type of cumulants,
known as ``partially  (time)-ordered'' cumulants \cite{VK,Deu,Mu1,Mu2,Shi}.
  This method was probably abandoned for two reasons. 
 First, it
incorrectly describes incoherent relaxation for $P(t)$ 
{\em for all values of } $\alpha$.  Secondly, it cannot
be obtained in a simple manner from the exact path integral expression. 
 The second objection is irrelevant, since
it is still possible that such an approximate resummation describes the exact behavior of $P(t)$ well.  The
first flaw, however, is quite serious.  Despite this, the expression obtained 
from the  ``partial ordering prescription'', or POP, which naturally resums to 
an exponential form, may be expected to give a better description of $P(t)$ in 
the incoherent region.  In fact, for values
of $\alpha$ greater than $\frac{1}{2}$, but not too large, 
this method indeed describes (nearly) exponential
relaxation.  Furthermore, as will be demonstrated in this paper,
 recent simulations of Egger and Mak \cite{EM} show that
 the POP method more accurately captures
the deep decay of $P(t)$  at zero temperature for $\alpha > \frac{1}{2}$ than does the COP (NIBA) method,
 {\em even before the algebraic behavior of the NIBA is manifested}.
 
It is well known that by choosing a particular ordering in a truncated cumulant expansion, 
we are implicitly assuming different
statistical properties for the relevant bath operators.  The first purpose of this paper is to specify
these statistical properties for the case of the spin-boson problem at zero temperature. 
 Using this 
``stochastic'' type intuition, we then discuss various cumulant ordering schemes and 
their associated descriptions 
of the behavior of $P(t)$ at $T=0$.  This paper is organized as follows: In Sec. 2 
we first present a new
derivation of the exact expression for $P(t)$ that allows for clear specification
 of the statistical properties
of the bath.  For this purpose, orthogonally to the conventional approach,
we first integrate out the spin degrees of freedom exactly.
In Sec. 3, we briefly discuss the COP and POP methods.  
We then turn to  a recently introduced new cumulant method, the
``non-crossing''  cumulants \cite{Spei,NSp1,NSp2}.  Lastly, in Sec. 4, 
we compare the methods to exact simulation result.

\section{Moment Expansion}

We begin with an explicit expression for $P(t)$ through fourth order in $\Delta$.  We could, if we wished,
obtain these terms from the exact path integral expression for $P(t)$, however, we believe that the method 
used in this section most clearly shows the connection to the stochastic methods 
upon which the cumulant expansions
are based.  In effect, our method offers another route to the formal expression of Ref. \cite{Legg,WB}.

We begin with the Hamiltonian (1) in the form,
\begin{eqnarray}\label{6}
H &=&H^{'} + \frac{\Delta}{2} \sigma_{x},\\
\label{7}
H^{'} &=& \sum_{k} \omega_{k} b_{k}^{\dagger} b_{k} - \sigma_{z} \sum_{k} g_{k}(b_{k}^{\dagger}+b_{k}) + \sum_{k} \frac{g_{k}^{2}}{\omega_{k}}.
\end{eqnarray}
The quantity we wish to calculate is $P(t)$ which is  defined as
\begin{equation}\label{8}
P(t) = \langle \sigma_{z}(t) \rangle = Z^{-1}  {\rm Tr}
 \left( \exp(i H t) \sigma_{z}(0)\exp(-i H t) \pi^{+} \exp(-\beta H^{'} )
\pi^{+}\right)
\end{equation}
where  
\begin{eqnarray}
Z &=& {\rm Tr} \left( \pi^{+} \exp(-\beta H^{'}) \pi^{+} \right), \nonumber\\
 \sigma_{z} &=& |L \rangle \langle L| - |R \rangle \langle R|, \nonumber\\
  \sigma_{x} &=&  |L \rangle \langle R|
+ |R \rangle \langle L|, \nonumber\\
\pi^{+} &=& \frac{1}{2}(1+\sigma_{z}),\nonumber
\end{eqnarray}
 and $\beta$ is the inverse temperature.
We now diagonalize (6) {\em in the spin manifold} with the use of a transformation
 employed by Shore and Sander \cite{SS,WAB}
in their study of the self-trapping of an exciton coupled to phonons, namely,
\begin{equation}\label{9}
U = \frac{1}{\sqrt{2}} \left( \begin{array}{cc} {1}& {-1} \\
                               {\phi}& {\phi} \end{array} \right),
\end{equation}
where \[ \phi = (-1)^{\sum_{k}b_{k}^{\dagger}b_{k}} = \exp \left( i \pi \sum_{k} b_{k}^{\dagger}b_{k} \right) . \]
The operator $\phi$ is seen to be the parity operator for the bath modes.  In the transformed picture,
we can express
\begin{equation}\label{10}
P(t) = - \widetilde{Z}^{-1} {\rm Tr} \left( \exp(i \widetilde{H} t ) 
\sigma_{x}(0) \exp( -i \widetilde{H} t) \widetilde{\pi}^{+} 
\exp(-\beta \widetilde{H}^{'} ) \widetilde{\pi}^{+} \right).
\end{equation}  
where
\begin{eqnarray}\label{11}
\widetilde{H} &=& \frac{\Delta}{2} \phi \sigma_{z} + \tilde{H}^{'},\\
\label{12}
\widetilde{H}^{'} &=& \sum_{k} \omega_{k} b_{k}^{\dagger}b_{k} + \sum_{k} g_{k}(b_{k}^{\dagger} + b_{k}),\\
\label{13}
\widetilde{\pi} &=& \frac{1}{2} (1-\sigma_{x}),
\end{eqnarray}
and $\widetilde{Z}$ is now defined with respect to $\widetilde{H}^{'}$ and $\widetilde{\pi}^{+}$.

We now perform the trace over the spin degrees of freedom in (10), leaving
\begin{equation}\label{14}
P(t) = {\rm Re} \left[ G(t)\right],
\end{equation}
where 
\begin{equation}
G(t) = {\rm Tr}_{b} \left( \exp(i H_{+} t ) \exp(-i H_{-} t)
\exp(-\beta \widetilde{H}^{'}) \right)\, / \, {\rm Tr}_{b}(\exp(-\beta \widetilde{H}^{'})),
\end{equation}
with \[H_{\pm} = \pm \frac{\Delta}{2} \phi + \tilde{H}^{'} .\]
This trace over the bath degrees of freedom is most easily performed in the small polaron representation,
defined by the transformation
\begin{eqnarray}\label{16}
U &=& \exp(\xi),\\
\label{17}
\xi &=& \sum_{k} \frac{g_{k}}{\omega_{k}} (b_{k} - b_{k}^{\dagger}).
\end{eqnarray}
In this picture, we may express $G(t)$ as
\begin{equation}\label{18}
\langle \exp_{\rightarrow}(i \int_{0}^{t} d\tau \eta(\tau) )\,
\exp_{\leftarrow}(i \int_{0}^{t} d\tau \eta(\tau) ) \rangle_{B},
\end{equation}
where
\begin{equation}\label{19}
\eta(t) = \frac{\Delta}{2} \exp[-\xi(t)]\,  \phi\,  \exp[\xi(t)],
\end{equation}
and
\begin{equation}\label{20}
\xi(t) =  \sum_{k} \frac{g_{k}}{\omega_{k}} (b_{k} e^{-i \omega_{k} t} - b_{k}^{\dagger} e^{i \omega_{k} t}).
\end{equation}
The averaging (denoted by $\langle ... \rangle_{B}$) is over the canonical 
ensemble of harmonic oscillators, 
(i.e., $\rho_{B} = \exp(-\beta \sum_{k} \omega_{k}b_{k}^{\dagger}b_{k})\, /\, {\rm Tr}_b
\exp(-\beta \sum_{k} \omega_{k}b_{k}^{\dagger}b_{k})$,
and $\exp_{\rightarrow}$ ($\exp_{\leftarrow}$) denotes a 
time ordered exponential with latest time to the right (left).   
From this point on, all averaging will be with respect to this ensemble, 
and we will drop the subscript ``$B$''.  Since the spin degree of freedom has been removed, 
our method allows us to focus on the bath operators that arise
in the expansion of $P(t)$.  Using the following properties of the parity operator, \[\phi\, \exp[\xi(t)] =
\exp[- \xi(t) ]\, \phi, \] and \[ \phi^{2} = 1, \]
we can show, through fourth order in $\Delta$, the moment expansion for $P(t)$,
\begin{equation}\label{21}
P(t) = 1 + \int_{0}^{t} dt_{1} \int_{0}^{t_{1}} dt_{2} \, m_{2}(t_{1} t_{2}) 
+ \int_{0}^{t}dt_{1} \int_{0}^{t_{1}}
dt_{2} \int_{0}^{t_{2}}dt_{3} \int_{0}^{t_{3}}dt_{4} \, m_{4}(t_{1},t_{2},t_{3},t_{4}) +... ,
\end{equation}
where the moments $m_{i}$ equal
\begin{eqnarray}\label{22}
m_{2}(t_{1},t_{2}) &=&  -\Delta^{2} {\rm Re}\, \langle B_{-}(t_{1}) B_{+}(t_{2}) \rangle,\\
m_{4}(t_{1},t_{2},t_{3},t_{4}) & = & \frac{\Delta^{4}}{4}\, {\rm Re}\,
\Big[ \langle B_{-}(t_{1})B_{+}(t_{2})B_{-}(t_{3})
B_{+}(t_{4}) \rangle  \nonumber \\
 &+&  \langle B_{-}(t_{2})B_{+}(t_{1})B_{-}(t_{3})B_{+}(t_{4}) \rangle \nonumber \\
 &+&  \langle B_{-}(t_{3})B_{+}(t_{1})B_{-}(t_{2})B_{+}(t_{4}) \rangle \nonumber \\
 &+&   \langle B_{-}(t_{4})B_{+}(t_{1})B_{-}(t_{2})B_{+}(t_{3}) \rangle\, \Big],
 \label{23}
\end{eqnarray}
and 
\begin{equation}\label{24}
B_{\pm}(t) = \exp[\pm2 \xi(t)].  
\end{equation}
Note that $m_{2n-1} = 0$.  
In this paper, we shall only use the first two nonvanishing moments, although it is a simple matter to execute
the expansion to an arbitrary order.  From (\ref{21})--(\ref{23}) we conclude that 
$P(t)$ is entirely determined by the statistical properties of the bath operators
$B_\pm(t)$ with respect to the canonical state of the bath.
Note that the operators $B_{\pm}$ always appear in pairs. In order to 
specify the statistics obeyed by the operators $B_{\pm}$, we now calculate the second and fourth
moment of the $B_{\pm}$'s. 
 It is a simple matter to show that 
\begin{equation}\label{25}
\langle B_{-}(t_{1})B_{+}(t_{2}) \rangle = \exp\left[-iQ_{1}(t_{1}-t_{2}) - Q_{2}(t_{1}-t_{2})\right],
\end{equation}
where
\begin{equation}\label{26}
Q_{1}(t_{1}-t_{2}) = 4 \sum_{k} \left( \frac{g_{k}}{\omega_{k}} \right)^{2} \sin (\omega_{k} (t_{1}-t_{2})),
\end{equation}
and
\begin{equation}\label{27}
Q_{2}(t_{1}-t_{2}) = 4 \sum_{k} \left( \frac{g_{k}}{\omega_{k}} \right)^{2} [1-\cos(\omega_{k}(t_{1}-t_{2}))]\coth(
 \beta \omega_{k}/2).
\end{equation}
Furthermore, by using the relation $e^{A+B} = e^A e^B e^{-{1\over2}[A,B]}$  and 
$\langle e^{\kappa  b^{\dagger}} e^{-\kappa^*b}\rangle = e^{-|\kappa|^2}$
for  $[A,B]$, $\kappa$ being  $c$-numbers,
one can also show that the operators $B_{\pm}$ have the following ``statistical'' property,
\begin{equation}\label{28}
\langle B_{-}(t_{1})B_{+}(t_{2})B_{-}(t_{3})B_{+}(t_{4}) \rangle = 
\frac{ \langle B_{-}(t_{1})B_{+}(t_{2}) \rangle
\langle B_{-}(t_{3})B_{+}(t_{4}) \rangle \langle B_{-}(t_{1})B_{+}(t_{4})
 \rangle \langle B_{-}(t_{2})B_{+}(t_{3}) \rangle}{
\langle B_{-}(t_{1})B_{+}(t_{3}) \rangle \langle B_{-}(t_{2})B_{+}(t_{4}) \rangle}.
\end{equation}
This property can be extended to an arbitrary number of $B_{\pm}$ pairs. 
 This gives a type of ``Wick''  theorem
for the operators $B_{\pm}$, and demonstrates the underlying reason why only 
the functions $Q_{1}$ and $Q_{2}$
appear in the exact path integral (see Eqs. (4-17) to (4-22) in Ref. \cite{Legg}).
 It can now be explicitly checked
that the expression (\ref{21}) is identical to the exact path integral expression, at least through the fourth
moment.  Note that the property (\ref{28}) is different than the statistical properties held by commonly used
stochastic processes such as Gaussian, two-state-jump, or Gaussian random matrix processes. 
 We will return to
this point in the next section.

The moment expansion itself is not a very useful scheme for describing dynamics, because an arbitrary 
truncation of the expansion leads to secular terms that grow with time.  We next resort to schemes that 
provide partial (approximate) resummations of the moment expansion to all orders.  Such schemes are the 
cumulant expansions that will be introduced in the next section.

\section{Cumulant Expansions}
We now discuss the various ordering prescriptions which allow for partial resummation of the expansion (21).
Each ordering method leads to a unique type of master equation \cite{NSp1}.
  We note that, when carried out to infinite
order, all of the ordering techniques give the same (exact) result. 
 When  truncated at a finite order, however,
the results are different.  In simple stochastic situations, when the
 generator for time evolution (the Liouville
operator) commutes with itself for all times, i.e., $[L(t),L(t^{'})] =0$, 
the use of a particular truncated cumulant expansion implies a
knowledge of the stochastic properties of the bath functions.  In simple cases, truncation of the cumulant 
expansion in the ``correct'' ordering prescription can lead to {\em exact} results that may be obtained
in the ``incorrect'' ordering prescription only at infinite order. 
 In the quantum case described by the Hamiltonian (1), where $[L(t),L(t^{'})] \neq 0$, truncation
of a cumulant expansion at finite order in any ordering prescription will never lead to exact results due to
the non-commutivity of the Liouvillian at different times 
\cite{Fox,Rips,Skin1}.   It is precisely this noncommutivity that leads to
the variety of time-orderings of the operators $B_{\pm}$ in the expression (\ref{23})
 for $m_{4}$.  Despite this fact, the statistical properties of the
bath operators still dictate the choice of an ordering prescription that provides the most rapid convergence of the
cumulant series (if such convergence exists) \cite{Skin2}.

We begin by discussing the chronological ordering prescription, or COP.  In this prescription, a master
equation of the form (see for instance \cite{Shi,NSp1})
\begin{equation}\label{29}
\frac{dP(t)}{dt} =\int_{0}^{t}  K^{COP}(t,\tau) P(\tau)\, d \tau
\end{equation}
is obtained.  The kernel $K^{COP}(t, \tau)$ is obtained from the moment expansion as 
\begin{equation}\label{30}
K^{COP}(t,\tau) = \gamma_{2}(t,\tau) + \sum_{n=2}^{\infty}
 \int_{0}^{\tau_{1}} d\tau_{2} ... \int_{0}^{\tau_{n-1}}
d\tau_{n}\,  \gamma_{n+1}(t,\tau_{1},...\tau_{n}).
\end{equation}
The COP cumulants, $\gamma$, are obtained from the moments by a recusion relation \cite{NSp1}.
In the present case this yields
\begin{eqnarray}
\gamma_{2n-1} & = & 0, \nonumber \\ 
   \gamma_{2}(t,\tau_{1}) & = &m_{2}(t,\tau_{1}), \nonumber \\
  \gamma_{4}(t,\tau_{1},\tau_{2},\tau_{3}) & = &
m_{4}(t,\tau_{1},\tau_{2},\tau_{3}) - m_{2}(t,\tau_{1}) \, m_{2}(\tau_{2},\tau_{3}), ..., .
\label{31}
\end{eqnarray}

For the simple case where the stochastic Liouvillian commutes with itself for all times, all
of the COP cumulants $\gamma_{n}$ vanish for $n \geq 3$ if the stochastic bath functions have the 
two-state-jump behavior \cite{Shi}
\begin{equation}\label{32}
\langle B(t_{1})B(t_{2})B(t_{3})B(t_{4}) \cdot \cdot \cdot \rangle = \langle B(t_{1})B(t_{2}) \rangle
\langle B(t_{3})B(t_{4}) \cdot \cdot \cdot \rangle,
\end{equation}
for $t_{1} > t_{2} > t_{3} > t_{4} \cdot \cdot \cdot $, where $B(t)$ is the stochastic bath function responsible
for system dissipation.  If these bath functions have different statistics, it may not be a good approximation 
to truncate the series at low orders.

Returning now to the quantum case of interest in this paper, 
we find at lowest order, as shown by Aslangul et. al. \cite{As1}, 
the NIBA equation for $P(t)$ 
\begin{equation}\label{33}
\frac{d P(t)}{dt} = \int_{0}^{t} m_{2}(t-\tau) P(\tau) d \tau,
\end{equation}
where, at $T =0$, using the ohmic constraint (\ref{3}), along with (\ref{22}), we may express
\begin{equation}\label{34}
m_{2}(t-\tau) = -\Delta^{2} Re \frac{1}{[1+ i \omega_{c}(t- \tau)]^{2 \alpha}}.
\end{equation}
As shown by Grabert and Weiss \cite{GW}, the solution to (\ref{33}) with the kernel (\ref{34})
 can be given for all $\alpha < 1$
(in the limit $\frac{\Delta}{\omega_{c}} \rightarrow 0$) by the Mittag-Leffler function \cite{Er},
\begin{equation} \label{35}
P_{NIBA}(y) = E_{2(1-\alpha)}(-y^{2(1-\alpha)}),
\end{equation}
where $y= \Delta_{\rm eff} t$ and 
\begin{equation}\label{36}
\Delta_{\rm eff} = \Delta \left[ \cos(\pi \alpha)
 \Gamma(1-2 \alpha) \right]^{\frac{1}{2(1-\alpha)}}
 \left( \frac{\Delta}{\omega_{c}}\right)^{\frac{\alpha}{1-\alpha}}.
\end{equation}
This solution shows damped oscillations for $\alpha < \frac{1}{2}$, and incoherent 
decay for $\alpha \geq \frac{1}{2}$.
This behavior has been qualitatively confirmed by Monte Carlo simulation \cite{EM}. 
 As mentioned in the introduction, the NIBA
cannot give the correct asymptotic decay of $P(t)$ (yielding 
 the algebraic decay $P(t) \propto t^{-2(1-\alpha)}$ rather than exponential decay), and is
unable to account for the depth of the decay in the region 
$\alpha \geq \frac{1}{2}$ even before  the incorrect
algebraic behavior sets in.  The NIBA is, however, known to work quite well for short times and weak coupling
strengths.  The analysis given in the last section provides a novel explanation for this fact.  For ``short''
times and ``small'' values of $\alpha$ the function $m_{2}(t)$
 is a rather broad, weakly decaying function of time.
When this is the case, the statistical property (\ref{28}) of the 
operators $B_{\pm}$ is approximately of the two-state-jump
form (\ref{32}) {\em as far as the integrations over the cumulants 
$\gamma_{n \geq 3}$ are concerned}.  This approximate
equivalence holds  in a {\it stochastic} sense, in that all of the four point
 correlation functions in $m_{4}$ (see
Eq. (\ref{23})) may be approximated by $m_{2}(t_{1},t_{2})\, m_{2}(t_{3},t_{4})$.  For such times and coupling
strengths, the NIBA will be essentially exact, as all COP cumulants for $n \geq 3$ 
will vanish when integrated.
We shall not provide precise meaning to the terms ``short'' or ``small'', 
although their meaning should be clear in the context of the present discussion, and could be quantified
without undue labor (in fact ``short'' and ``small'' will be 
coupled in the sense that the effective timescale of
oscillation or decay, $(\Delta_{\rm eff})^{-1}$, depends on $\alpha$). 
Note that the statistical property (\ref{28})
trivially gives two-state-jump behavior for $\alpha =0$, which leads to the correct 
behavior $P(t)= \cos(\Delta t)$.
While this is obvious, other cumulant techniques (such as those discussed below) do not embody this type of
statistics for $\alpha =0$, and cannot give the correct, freely oscillating solution for zero
coupling strength upon truncation at second order.
   The statement (often given in the literature \cite{Stock}) that NIBA works 
for weak coupling because it is a perturbative
scheme is thus not strictly correct.

The (somewhat heuristically) demonstrated fact that the property (\ref{28})
 can resemble two-state-jump behavior
under certain circumstances leads one to believe that extending the COP scheme to fourth order would not
be useful, since this property is reflected in the vanishing of all COP cumulants higher than the second.
Extending the COP method to fourth order {\em does not give a method for computing ``interblip'' interactions}
in the language of Ref. \cite{Legg}.

We now turn to the partial ordering prescription, or POP.  At second order, this method was applied by 
Aslangul et. al. \cite{As2} to the spin-boson problem at $T =0$. 
 The POP master equation has a convolutionless \cite{Shi,NSp1}
form
\begin{equation}\label{37}
\frac{dP(t)}{dt} = \left( \int_{0}^{t} K^{POP}(\tau) d \tau  \right) P(t).
\end{equation}
$K^{POP}(t)$ may be obtained from the moments 
\begin{equation}\label{38}
K^{POP}(t)  = \sum_{n=1}^{\infty} \int_{0}^{t} d \tau_{1} \int_{0}^{\tau_{1}} d\tau_{2} 
... \int_{0}^{\tau_{n-1}}
d \tau_{n} \, \theta_{n+1}(t,\tau_{1},...\tau_{n}),
\end{equation}
where
\begin{eqnarray}
\theta_{2n-1} & = & 0, \nonumber \\
  \theta_{2}(t,\tau_{1}) & = &  m_{2}(t,\tau_{1}), \nonumber \\
  \theta_{4}(t,\tau_{1},\tau_{2},\tau_{3}) & = &
m_{4}(t,\tau_{1},\tau_{2},\tau_{3}) - m_{2}(t,\tau_{1}) \, m_{2}(\tau_{2},\tau_{3}) \nonumber \\ 
 &-& m_{2}(t,\tau_{2}) \, m_{2}(\tau_{1},\tau_{3})
- m_{2}(t,\tau_{3}) \, m_{2}(\tau_{1},\tau_{2}), ...,
\label{39}
\end{eqnarray}

The POP resummation is exact at second order for the simple case of a classical Gaussian stochastic process.
We note that the statistical property (\ref{28}) appears to be very different from the standard Wick theorem for
Gaussian processes.  We may still expect that the POP method is better suited for the incoherent regime
$\alpha \geq \frac{1}{2}$ for the following reasons.  First, the POP technique resums to an exponential
form, which is expected to better capture the long time behavior of $P(t)$.  In general, the POP method
sums (infinitely) more terms than the COP method does.  For example, expansion of the second order
truncation in the COP gives, to fourth order
\[ P(t) = 1+\int_{0}^{t} d t_{1} \int_{0}^{t_{1}} d t_{2}\, m_{2}(t_{1},t_{2}) + \int_{0}^{t} d t_{1}
\int_{0}^{t_{1}} dt_{2} \int_{0}^{t_{2}} dt_{3} 
\int_{0}^{t_{3}} dt_{4} \, m_{2}(t_{1},t_{2}) \, m_{2}(t_{3},t_{4}) + ...
 \]
whereas the POP gives 
\begin{eqnarray}
 P(t) & = & 1 + \int_{0}^{t} dt_{1} \int_{0}^{t_{1}} dt_{2} \, m_{2} (t_{1},t_{2})  + 
 \int_{0}^{t} dt_{1} \int_{0}^{t_{1}} d t_{2}    \int_{0}^{t_{2}} dt_{3} \int_{0}^{t_{3}} dt_{4}\,
    m_{2}(t_{1},t_{2}) \, m_{2}(t_{3},t_{4}) \nonumber \\
 &+&   m_{2}(t_{1},t_{3}) \, m_{2}(t_{2},t_{4}) + m_{2}(t_{1},t_{4}) \, m_{2}(t_{2},t_{3})  + ...  \ .\nonumber
\end{eqnarray}  
Clearly, the extra terms do not insure a more accurate result.  For example, for weak coupling strengths,
the POP must be carried out to infinite order to obtain coherent behavior.  However, in the incoherent
regime, the effective timescale, defined by (\ref{36}) 
is very long, while the decay of the function $m_{2}(t)$
is ``slow''  (algebraic).  In this case, we may expect that we
 must include terms like $m_{2}(t_{1},t_{4}) \, m_{2}(t_{2},t_{3})$
that extend over large portions of the integration region.  As we will show in the next section, 
the POP method captures the behavior of $P(t)$ better than the COP method in the incoherent regime, even
before the full asymptotic time behavior is displayed.

At second order, ($K^{POP}(t,\tau)=m_{2}(t-\tau)$) the POP equation (38) may be solved \cite{As2},
\begin{equation}\label{40}
P(t) = \exp\left[ \frac{\Delta^{2}}{4 \omega_{c}^{2}} \frac{1}{(\alpha-1/2)(1-\alpha)}
 \left( 1-\frac{\cos(2(1-\alpha)
\tan^{-1}\omega_{c} t)}{(1+\omega_{c}^{2} t^{2})^{\alpha -1}} \right) \right].
\end{equation}
Note that equation (\ref{40}) describes a {\it stretched exponential}
 rather than exponential decay.  For values of $\alpha$
that are not too much larger than $\frac{1}{2}$, however, the POP expression should be much more accurate
than the COP expression, at least at  second order.

We have now given some motivation for the belief that the COP method (at lowest order) should give a better description of 
$P(t)$  in the region $\alpha < \frac{1}{2}$  while the POP method should be better in the incoherent region $\alpha \geq \frac{1}{2}$.
We now ask whether there is a summation method that is a ``hybrid'' of the two methods, in the sense that it can incorporate at low order
features of the COP and POP methods.  In the theory of stochastic processes, 
such a technique has recently been developed \cite{Spei,NSp1,NSp2}.
  This method is based on the summation
of ``non-crossing'' (NC) cumulants (for a precise definition see Refs. \cite{Spei,NSp1,NSp2}).  
For simple  stochastic situations, if the coupling is not too strong, the NC
technique (including terms up to fourth order) has been shown 
to interpolate between the two-state-jump behavior and the Gaussian behavior \cite{NSp1}.

The NC description
leads to a {\it nonlinear}
 equation of motion for $P(t)$ \cite{NSp1}, which at second  order, may be expressed
\begin{equation}\label{41}
\frac{dP(t)}{dt} = M(t),
\end{equation}
where
\begin{equation}\label{42}
M(t) =  \int_{0}^{t} dt_{1}\,  \zeta_{2}(t-t_{1}) P(t-t_{1})P(t_{1}).
\end{equation}
To fourth order the master equation for $P(t)$ in the NC scheme reads 
\begin{equation}\label{43}
\frac{dP(t)}{dt} = M(t) + \int_{0}^{t}dt_{1} \int_{0}^{t_{1}} dt_{2} \int_{0}^{t_{2}}dt_{3} 
\, \zeta_{4}(t,t_{1},t_{2},t_{3})P(t-t_{1})P(t_{1}-t_{2})P(t_{2}-t_{3})P(t_{3}).
\end{equation}
As in the previous two case, the NC cumulants, $\zeta$,  may be obtained from the moments by
a recursion relation \cite{Spei,NSp2}. In the present case this yield up to fourth order
\begin{eqnarray}
\zeta_{2n-1} & = & 0, \nonumber \\ 
   \zeta_{2}(t,t_{1})  & = & m_{2}(t,t_{1}), \nonumber \\ 
   \zeta_{4}(t,t_{1},t_{2},t_{3})  & = & m_{4}(t,t_{1},t_{2},t_{3}) - m_{2}(t,t_{1}) \, 
   m_{2}(t_{2},t_{3})
- m_{2}(t,t_{3}) \, m_{2}(t_{1},t_{2}), ... \ .
\label{44}
\end{eqnarray}
It is clear that in appearance, the NC cumulants are a ``compromise'' between the COP and POP cumulants. 
 We note in passing two interesting facts.  First, 
in the stochastic realm, the NC ordering prescription truncated at second order 
is exact for the case of a stochastic bath modeled by symmetric $(N \times N)$ Gaussian random matrices
for the $B$'s.
  In this case, the ``crossing contraction'' $m_{2}(t,t_{2}) \, m_{2}(t_{1},t_{3})$ 
  vanishes by means of a $\frac{1}{N}$ argument for $N \rightarrow \infty$.
   This leads naturally to the equation (\ref{42}) 
   and its systematic generalization (\ref{43}) through the NC cumulants. Eq. (\ref{42})
   has first been derived by 
   Kraichnan \cite{Krai1,FB,Krai2,Krai3} in the field of turbulences and 
   fluid dynamics.  Our motivation for the application
of this method is {\em not} based on a stochastic type of reasoning, 
but on the fact that in simple situations this ordering prescription may combine the benefits of 
the COP and POP methods.

Before concluding this section, we would like to 
apply all three ordering prescriptions to the case $T=0$, $\alpha=\frac{1}{2}$.  
Here, it is known that the ``exact'' (in
the sense specified in Ref. \cite{Legg}) result for $P(t)$ is 
\begin{equation}\label{45}
P(t) = \exp\left(-\frac{\pi}{2} \omega_{c} \left( \frac{\Delta}{\omega_{c}} \right)^{2} \right)
\end{equation}
in the limit $\frac{\Delta}{\omega_{c}} \rightarrow 0$.  
Note that in this limit, the second moment  becomes 
$\delta$--correlated $(t\ge 0)$
\[ m_{2}(t) \rightarrow \frac{\pi \Delta^{2}}{2 \omega_{c}} \delta(t). \] 
 Using the fact that $P(0) =1$, it is clear that 
 { \em all three ordering prescriptions give the same result} given by (\ref{45})
at second order.  
 Hence, the value $\alpha = {1\over 2}$ corresponds to the {\it white noise} limit
 of the bath operators $B_\pm(t)$.

\section{Results and Conclusions}

Before comparing the results of the three ordering methods, we make some comments 
on the methods discussed in Sec. 3.  We have shown how three different 
cumulant methods give rise to different master equations with different properties. 
 We have tried to physically  motivate when each approach should have  success when 
 applied to the spin-boson problem at zero temperature.  Note that in general, the 
 discussion of convergence of each cumulant series is a difficult task.  This task is 
 made more difficult by the fact that, at zero temperature,
the algebraic decays of the bath correlation functions leave us with no clearly 
defined relaxation time for the bath.  This means that we will rely almost exclusively
on physical considerations and comparison with accepted results to determine the 
success or failure of the methods employed.  The case of finite temperature, which can be studied
by the same methods employed here, is often easier in this respect. 
 If an exponential correlation time $\tau_{b}$ can be assigned to the decaying 
 bath correlation functions, than it is possible to consider a systematic 
 expansion in $ \Delta_{\rm eff} \tau_{b}$ provided that this dimensionless parameter is small.  
 When this is the case, the POP  provides the most facile way of  systematically summing terms  
  in the parameter $\Delta_{\rm eff} \tau_{b}$ \cite{VK}. In case of ohmic dissipation
 and finite temperature $T$, the characteristic correlation time of the $B_\pm(t)$
 is given by $\tau_b \sim (2\pi\alpha T)^{-1}$ \cite{Legg}. This point of view
provides a novel explanation for the 
 familiar statement that the NIBA works well in the incoherent tunneling
 regime  $ \Delta_{\rm eff} \tau_{b}\ll 1$.
 In a stochastic language, this parameter region corresponds to the {\it narrowing} or
 {\it Markov} limit of the $B_\pm$'s. Similarly to the white noise limit mentioned
 above, one finds that all three cumulants schemes work well already at second order and provide essentially
 the same behavior, $P(t)\approx \exp[-\Delta_{\rm eff}^2(T)\tau_b]$
with $\Delta_{\rm eff}(T)\propto T^\alpha$ \cite{Legg}.

Since we expect the NIBA to be accurate for very weak coupling strengths, 
we first turn to the case of weak to intermediate coupling strength, $\alpha = 0.3$.
For coupling strengths in this range, simulations at low temperatures have shown 
that the NIBA is qualitatively correct in predicting damped oscillations,
but may fail in predicting the damping strength.
  An example of this is given by the simulations of Makarov and Makri \cite{Makri1}
   which show that for intermediate coupling, NIBA may fail by 
     slightly underestimating the number of oscillations in $P(t)$.
       We note, however, that these simulations were carried out for values of 
       $\frac{\Delta}{\omega_{c}}$ that are not very small.  In Fig.1, 
       we plot the NIBA (second order COP) solution for $P(t)$ against 
       the solutions obtained from second and fourth order truncations of
 the non-crossing cumulant method, and the second order POP.  Note that, as 
expected, the second order POP solution for $P(t)$ fails to produce any oscillations. 
 We expect that for $\alpha < \frac{1}{2}$ the POP will always be inaccurate at low orders.
  The second order non-crossing cumulant solution for $P(t)$, obtained from the Kraichnan-type equation (\ref{41})--(\ref{42}) 
is similar to the NIBA solution, although the oscillation in $P(t)$ is much weaker.  The fourth order noncrossing 
cumulant solution gives a first oscillation which is very similar in magnitude 
to the NIBA solution, however, it describes one extra weak remnant of an oscillation. 
 This behavior is very similar to the behavior displayed in the exact simulations of Makarov and Makri
 \cite{Makri1}.
   Although this example represents only one value of $\alpha$, similar 
   results may be obtained for all moderately strong values of $\alpha$ up 
   to $\alpha = \frac{1}{2}$.  Thus, it appears that the non-crossing 
   scheme works well in incorporating the qualities of the COP method 
   for moderate values of $\alpha$ when $\alpha < \frac{1}{2}$.

One may be a bit alarmed by the magnitude of the difference between the second and 
fourth order plots of $P(t)$ obtained via the non-crossing cumulant method.
This is not necessarily indicative of a lack of convergence in the summation of 
the cumulants.  As an example to support this, we mention the cumulant expansion results 
of Aihara, Budimir, and Skinner \cite{Skin2}.
  For a different model of relaxation, these authors compared the results 
  obtained from truncation of the POP expansion at sixth order with exact 
  simulations.  For coupling strength that were not too large, they found 
  that by sixth order, the cumulant method quantitatively describes the 
  results obtained from simulations.  In cases where convergence was obtained, 
  there was a large difference between the solutions obtained from truncating 
  the cumulant series at second and fourth orders.  In such cases, the fourth 
  order solution was nearly quantitative, but sightly {\em overestimated} the 
  exact result.  We believe that this is precisely what is occurring here, 
  although we have no direct evidence (i.e., a sixth order result) for this belief. 
   We base this statement on the similarity
 with exact simulations at slightly different coupling strengths, 
the fact that the fourth order non-crossing result shifts $P(t)$ 
in the right direction, and the reasonable appearance of the result.

We now turn to the relaxation of $P(t)$ in the incoherent regime $\alpha \geq \frac{1}{2}$. 
 Here, the beautiful path integral simulations of Egger and Mak \cite{EM}
  provide a means of comparing the cumulant expansion methods with exact results. 
   In this region, we expect the POP to be most successful, while the NIBA (second order COP) 
   is expected to be worse.  Based on experience with simple stochastic situations, we hope, 
   as in the coherent portion of the coupling space, that the non-crossing scheme can capture 
   the essence of the POP in this regime.  In Fig. 2, we show the decay of $P(t)$ calculated 
   by differing ordering prescriptions for $\alpha = 0.6$.  It must be noted that 
   the simulations were carried out for long times, but not long enough to show 
   the asymptotic algebraic decay of the NIBA (second order COP) solution of $P(t)$, 
   or the asymptotic exponential decay of the exact solution.  Regardless, the second 
   order truncation of the POP still gives the most accurate description 
of the decay of $P(t)$.  As we had hoped, the second order non-crossing technique is 
nearly identical to the POP for this value of $\alpha$.  Figure 3 shows the results for $\alpha = 0.7$.
  Again, the POP performs the best, while the second order non-crossing scheme over estimates the decay. 
   As in the case of weaker coupling, we see if truncation after fourth order 
   in the non-crossing cumulants can properly correct the second order result.  
   This test is shown in Fig. 4.  While the results appear to show that the 
   non-crossing scheme is converging to a POP-like description of $P(t)$, 
   we again must exercise caution due to the lack of further concrete evidence for this belief.
For such large value of $\alpha$, it is quite possible that the cumulant methods break down.

One interesting property displayed in Fig. 3 and Fig. 4 is the near
 {\em quantitative} agreement between the POP description of $P(t)$ 
 and the exact simulation of $P(t)$ for moderately long times. 
  In order to test if this is a coincidence, we have computed
   $\int_{0}^{y} d y^{'} \int_{0}^{y^{'}} dy^{''}K^{POP}_{2} (y^{''})$ and
     $\int_{0}^{y} d y^{'} \int_{0}^{y^{'}} dy^{''}K^{POP}_{4}(y^{''})$
where $K_n^{POP}(t)$ is the $n$-th term in the expansion (\ref{38}). 
      If the integrated second order POP cumulant is of order one for a
       given time interval, while all other POP cumulants are small 
       when integrated over the appropriate time domain, then we 
       expect the truncation at second order to be a good approximation.
         While we cannot study all the POP cumulants, we have studied 
         the second and the fourth.  In Fig. 5, we compare the properties 
         of the second and fourth POP cumulants for $\alpha = 0.7$. 
          For $y = 1.4$ to $y = 2$ (the boundary of the simulation results of Egger and Mak \cite{EM}),
           we see that the contribution from the second POP cumulant is at least
 ten times greater than the contribution from the fourth cumulant.  
 This strongly suggests that the agreement of the second order POP 
 method with the exact simulations is no coincidence.  In fact, 
 the agreement between (41) and the simulation occurs precisely 
 in the interval where the second order cumulant dominates the 
 fourth order cumulant.  Since the slopes of the two curves 
 suggest that this behavior continues for some time, we feel 
 there is strong evidence for the somewhat remarkable conclusion that,
  for significant intermediate times, the decay of $P(t)$ is 
  quantitatively described by a stretched exponential. 
   For longer times, the decay is most likely purely exponential.

We now summarize the results presented in this paper. 
 We first carried out a new derivation of the moment expansion for the variable 
 $P(t)$ in the spin-boson problem.  We then used this derivation to discuss the 
 ``statistical'' properties of the relevant bath operators.  Using the moment 
 expansion, we  first discussed the chronological and partial ordering prescriptions 
 that involve different types of cumulants.  We discussed the merits and drawbacks of each method. 
  In an effort to combine the merits of the COP and the POP, we applied the non-crossing scheme.
    Specializing to the case of zero temperature, we tested each method, including 
    fourth order terms when necessary.  Our results show that the non-crossing scheme is a 
    promising candidate for combining the virtues of the COP and POP, especially for
     intermediate values of $\alpha$ on either side of the coherent-incoherent
      transition value of $\alpha=\frac{1}{2}$. We note that more work should 
      be done to test the validity
of this claim.  Lastly, we have provided evidence to support the belief that
 the stretched exponential behavior described by second order truncation of 
 the POP in the incoherent portion of coupling space may infact be very 
 accurate for intermediate times. 

\section*{Acknowledgments}

We would like to thank the NSF for partial support of this research. 
 D. R. R. would like to thank the Air Force Office of Scientific 
 Research for financial support,  F. L. H. B. would like to thank the NSF for financial support,
and P. N. would like to thank the 
 Alexander von Humboldt foundation for financial support.
    We would also like to thank Professor C. Mak for providing the 
    data from his simulation results, and Professors R. Silbey and R. Speicher for 
    useful discussions.

\newpage

\noindent {\bf Figure Captions}

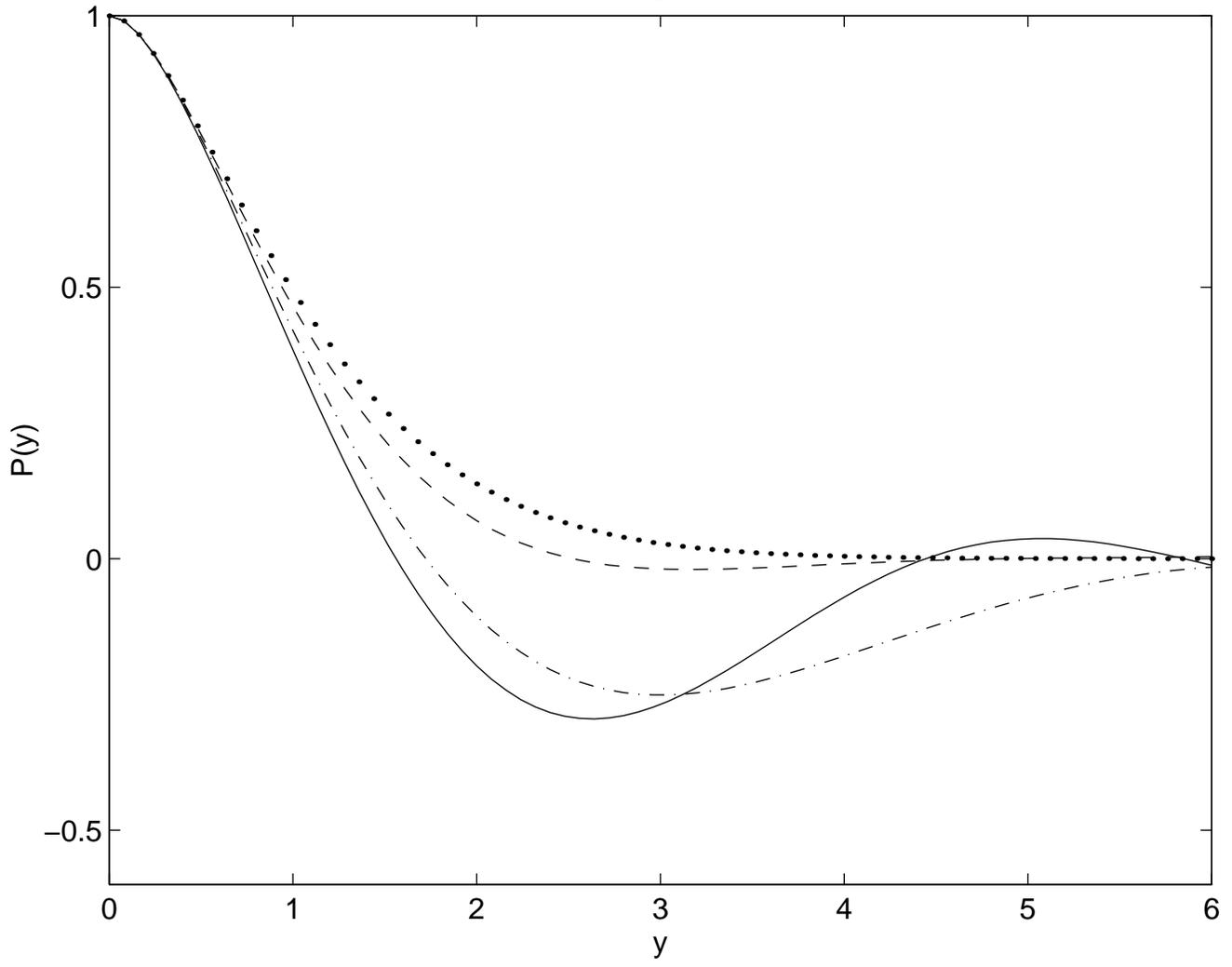
\noindent \underline{Fig.1:} Zero temperature plot of $P(y)$ ($y = \Delta_{\rm eff} t$) for   $\alpha = 0.3$ and $\frac{\omega_{c}}{\Delta} = 6$. 
 The dotted line is the second order POP result, the dashed line is the second 
order non-crossing cumulant result, the dash-dotted line is the NIBA (second order COP) result, 
and the solid line is the fourth order non-crossing cumulant result.

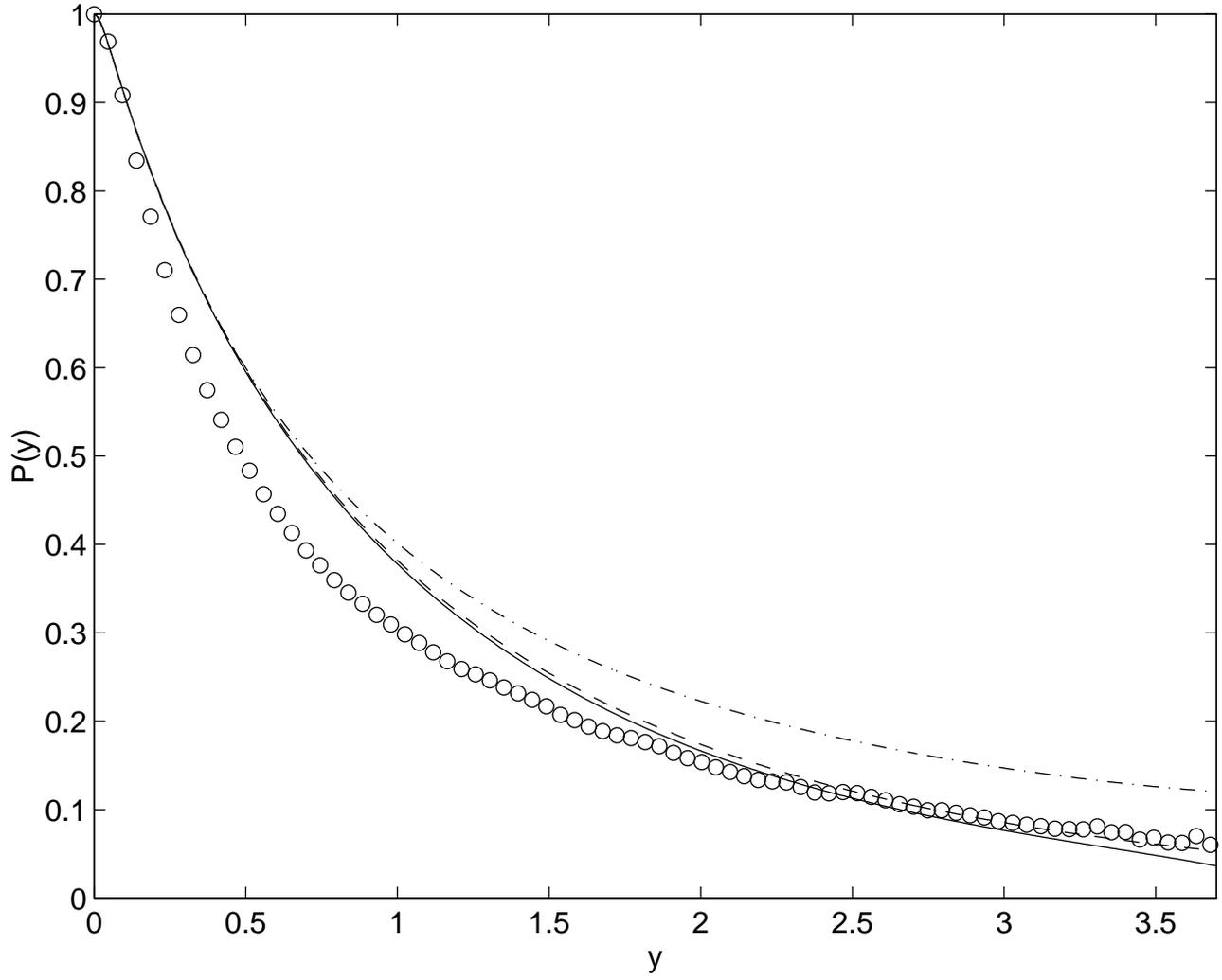
\noindent \underline{Fig.2:} Zero temperature plot of $P(y)$ ($y = \Delta_{\rm eff} t$) for   $\alpha = 0.6$ and $\frac{\omega_{c}}{\Delta} = 6$.
  The dash-dotted line is the NIBA (second order COP) result, the dashed line
is the second order POP result, the solid line is the second order non-crossing cumulant result, and the open circles are  the simulation 
result of Egger and Mak \cite{EM}.

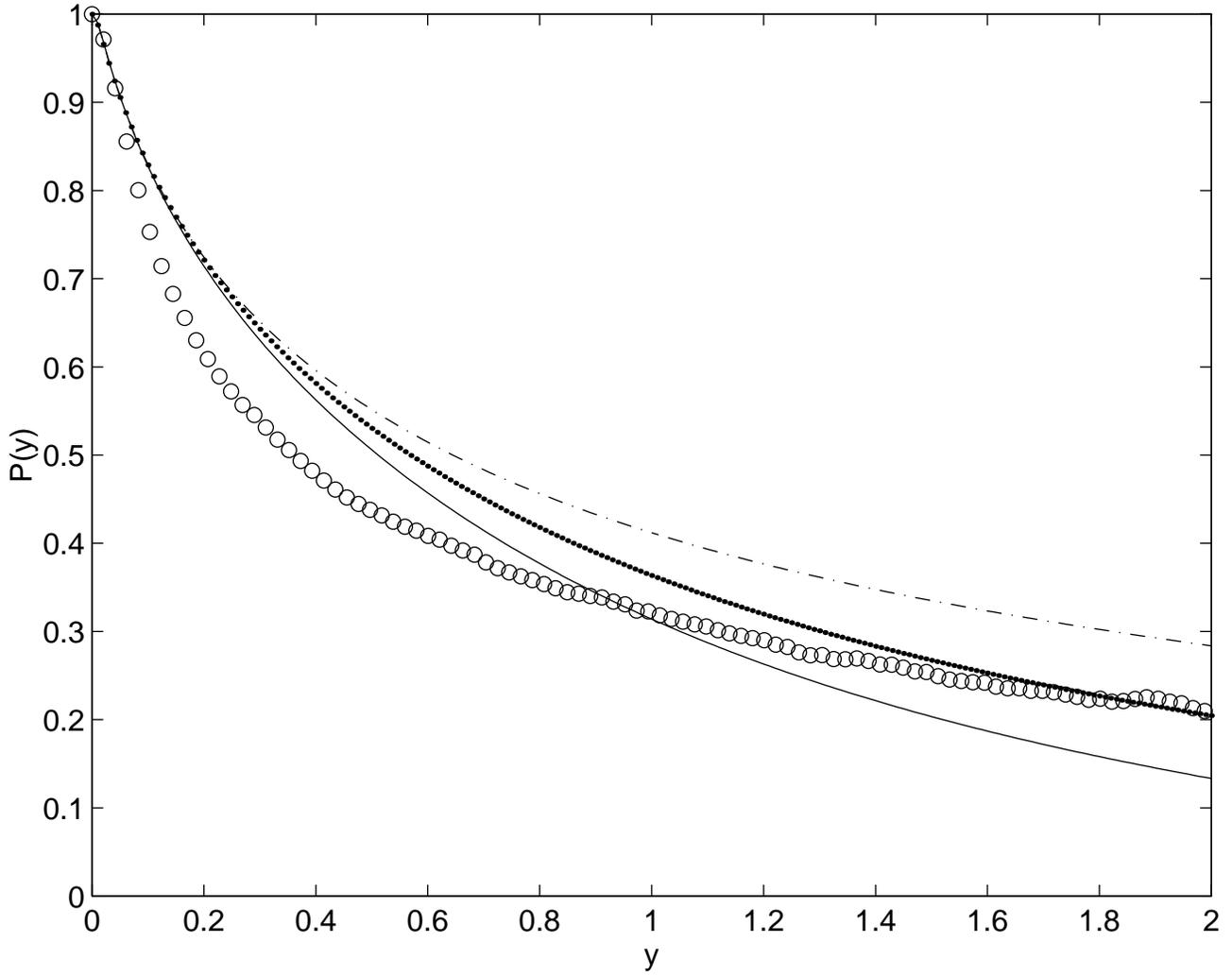
\noindent \underline{Fig.3:} Zero temperature plot of $P(y)$ ($y = \Delta_{\rm eff} t$) for   $\alpha = 0.7$ and $\frac{\omega_{c}}{\Delta} = 6$. 
The dash-dotted line is the NIBA (second order COP) result, the dotted line is 
the second order POP result, the solid line is the second order non-crossing cumulant result, and the open circles are 
 the simulation result of Egger and Mak \cite{EM}.

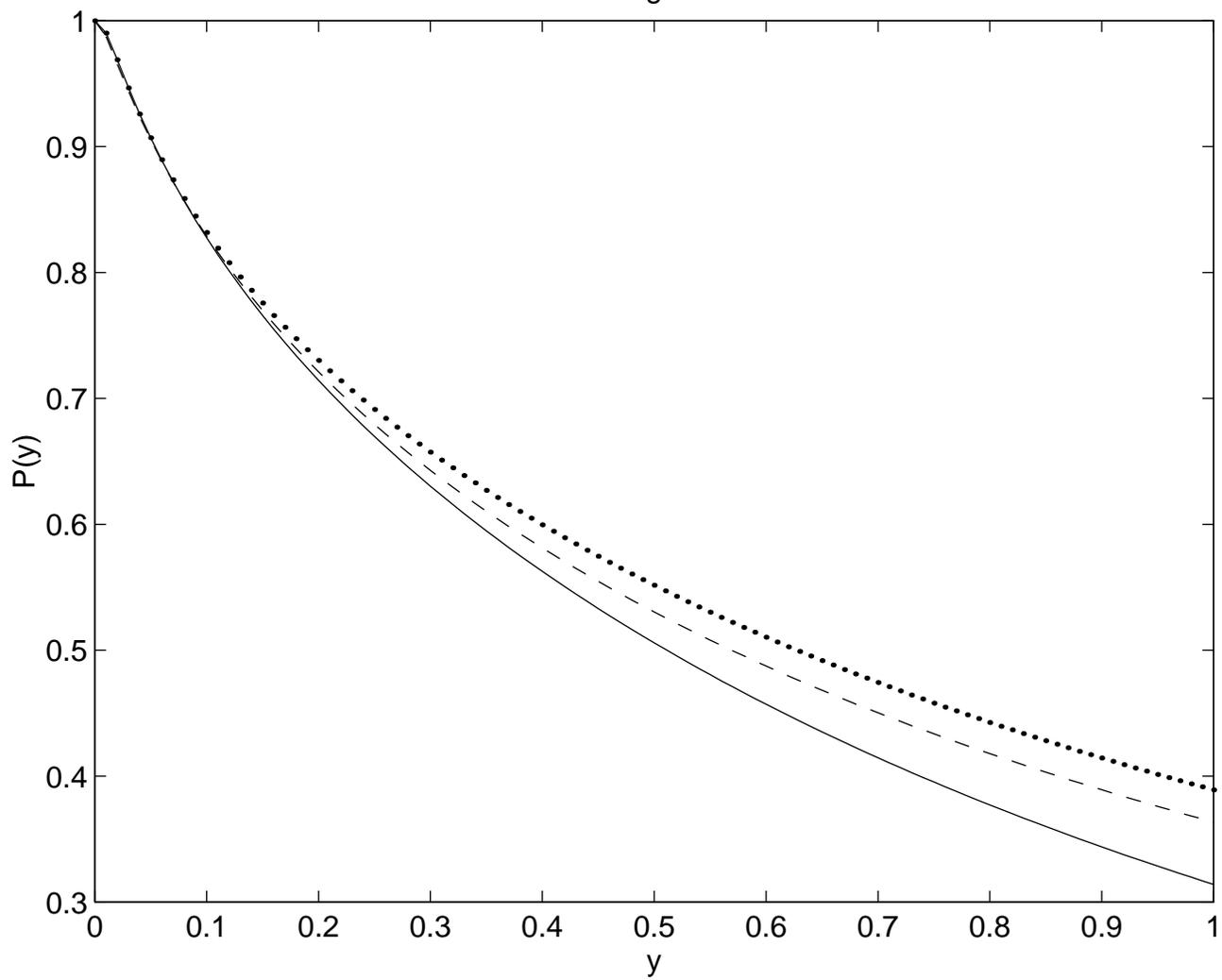
\noindent \underline{Fig.4:} Zero temperature plot of $P(y)$ ($y = \Delta_{\rm eff} t$) for   $\alpha = 0.7$ and $\frac{\omega_{c}}{\Delta} = 6$. 
The dotted line is the fourth order non-crossing  cumulant result, the dashed line is the second order POP result, 
and the solid line is the second order non-crossing cumulant result.  Note the change in the $x$-axis.

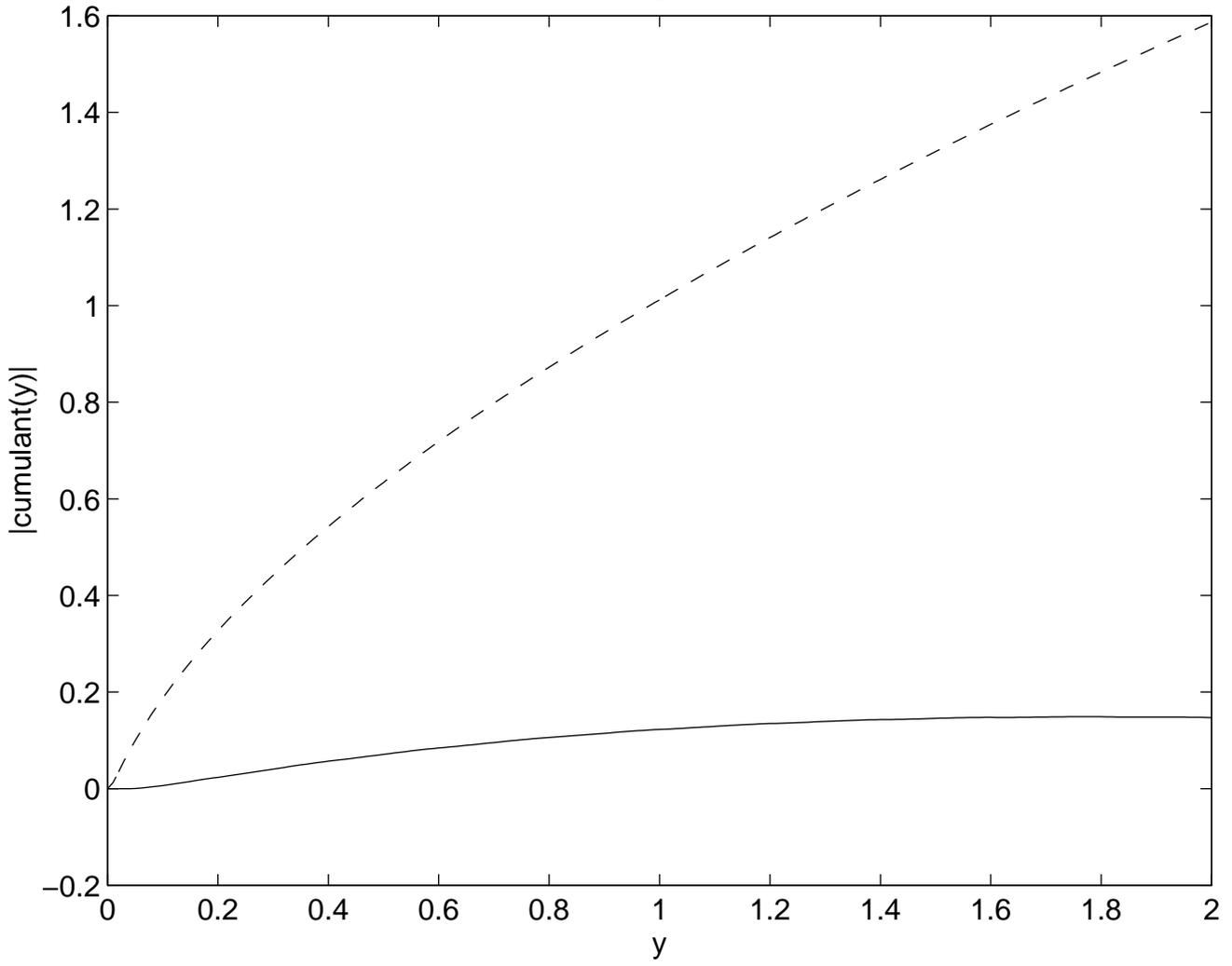
\noindent \underline{Fig.5:}  Relative magnitude of second and fourth cumulant effects in the POP for $\alpha = 0.7$. 
 The dashed line shows $|\int_{0}^{y} dy^{'} \int_{0}^{y^{'}} dy^{''} K^{POP}_{2}(y^{''})|$  and the solid line shows 
$|\int_{0}^{y} dy^{'} \int_{0}^{y^{'}} dy^{''} K^{POP}_{4}(y^{''})|$.  $K^{POP}(t)$ is defined in Eq. (38).

\end{document}